\documentclass[letterpaper,journal]{IEEEtran}
\usepackage{graphicx}
\usepackage{float}
\usepackage{amsmath}
\usepackage{epstopdf}
\usepackage{cases}
\usepackage{color}
\usepackage{xcolor}
\usepackage{algorithm}
\usepackage{algpseudocode}
\usepackage{amsmath}
\usepackage{amssymb}
\usepackage{subcaption}
\usepackage{soul}
\usepackage{hyperref}
\makeatletter
\renewcommand{\maketag@@@}[1]{\hbox{\m@th\normalsize\normalfont#1}}%
\makeatother
\usepackage{stfloats}
\usepackage{cite}
\usepackage{makecell}
\usepackage{multirow}
\usepackage{ulem}

\usepackage{tabularx}  % 自适应列宽
\usepackage{booktabs}  % 三线表样式

\usepackage[table]{xcolor} 

\ifCLASSINFOpdf
\else
\fi

\usepackage{mathtools}

\usepackage{acro}[=v3]
\acsetup{
	use-id-as-short=true,
	make-links=false,
	link-only-first=true,
}
\makeatletter
\DeclareAcronym{AI}{
	long={artificial intelligence},
}
\DeclareAcronym{LEO}{
	long={low Earth orbit},
}
\DeclareAcronym{6G}{
	long={The sixth-generation},
}
\DeclareAcronym{BS}{
	long={base stations},
}
\DeclareAcronym{3GPP}{
	long={the 3rd generation partnership project},
}
\DeclareAcronym{NTN}{
	long={non-terrestrial network},
}
\DeclareAcronym{QoS}{
	long={quality of service},
}
\DeclareAcronym{QoE}{
	long={quality of experience},
}
\DeclareAcronym{TSN}{
	long={time-sensitive networking},
}
\DeclareAcronym{MIMO}{
	long={multiple-input multiple-output},
}
\DeclareAcronym{gNB}{
	long={next generation node B},
}
\DeclareAcronym{UE}{
	long={user equipment},
}
\DeclareAcronym{DNN}{
	long={deep neural network},
}
\DeclareAcronym{SWaP}{
	long={size, weight and power},
}
\DeclareAcronym{EO}{
	long={Earth observation},
}
\DeclareAcronym{OOD}{
	long={out-of-distribution},
}
\DeclareAcronym{FL}{
	long={federated learning},
}
\DeclareAcronym{COR}{
	long={critical observation recall},
}
\DeclareAcronym{SMU}{
	long={sustainable mission utility},
}
\DeclareAcronym{AUROC}{
	long={area under the receiver operating characteristic curve},
}
\DeclareAcronym{ISL}{
	long={inter-satellite link},
}
\DeclareAcronym{LLM}{
	long={large language model},
}
\DeclareAcronym{VLM}{
	long={vision-language model},
}
\DeclareAcronym{LCM}{
	long={lifecycle management},
}

\immediate\write18{texcount -inc -sum main.tex > wordcount.txt}

\begin{document}
\title{Orbit-AWARE: From Onboard AI Deployment \\ to Sustainable In-Orbit Intelligence}
\author{
        Yunxin Geng,~\IEEEmembership{Graduate Student Member,~IEEE,}
        Kaige Zhu, Tongtong Shu, Zhan Gao, 
        Yanxiao Li, Yan Liu, Weiwei Jiang,~\IEEEmembership{Senior Member,~IEEE,} Zhiqing Wei,~\IEEEmembership{Member,~IEEE,} and Zhiyong Feng,~\IEEEmembership{Senior Member,~IEEE}
        
\thanks{
Yunxin Geng, Kaige Zhu, Zhan Gao and Yan Liu are with State Key Laboratory of Satellite Network, Shanghai Satellite Network Research Institute Co., Ltd., and Shanghai Key Laboratory of Satellite Network, Shanghai, China.
Yunxin Geng is also with Beijing University of Posts and Telecommunications, Beijing, China.

Tongtong Shu and Yanxiao Li are with China Satellite Network Group Co., Ltd., Beijing, China.
Yanxiao Li is also with Beijing University of Posts and Telecommunications, Beijing, China.

Weiwei Jiang, Zhiqing Wei and Zhiyong Feng are with Beijing University of Posts and Telecommunications, China.
Weiwei Jiang is also with Anhui Provincial Key Laboratory of Multimodal Cognitive Computation, Anhui University, Hefei, China.
}
\thanks{Corresponding authors: Yan Liu (\href{mailto: }{liu-y15@tsinghua.org.cn}), Weiwei Jiang (\href{mailto: jww@bupt.edu.cn}{jww@bupt.edu.cn}).}
}
\maketitle

\begin{abstract} 
The growing role of space computing in time-critical satellite services makes the long-term reliability of onboard artificial intelligence (AI) a mission-level requirement.
Model drift can undermine this reliability through critical-observation suppression or noncritical priority flooding, displacing both mission information and evidence for model correction.
To address this drift-induced evidence displacement, this article presents Orbit-AWARE, a safeguard-first lifecycle management (LCM) architecture.
A comparison of existing approaches motivates cooperative Data, Cognitive, and Trust Planes for continuous monitoring, event-triggered agentic orchestration, and deterministic governance.
The architecture retains a model-independent sample stream before degradation is confirmed and coordinates evidence selection and exchange for diagnosis and ground-assisted adaptation.
Candidate updates are verified before control over data disposition is restored.
In an xBD-based emergency Earth observation (EO) case study, Orbit-AWARE delivers 14.3\% more critical observations and achieves 8.3\% higher sustainable mission utility than the Rule-Based Drift Loop under a common 10\% downlink budget.
Finally, we discuss future research directions for extending LCM through task-aware reliability assessment, memory- and tool-enabled onboard agents, and evidence-centric space networking.
This work makes evidence availability and recovery authorization explicit design objectives for dependable onboard AI services.
\end{abstract}

\section{Introduction}
% 技术发展—具体风险—研究缺口—生命周期方案—本文贡献
\label{sec:introduction}
% 1. 从卫星技术发展收束到星载AI的数据处置权
% 概括低轨网络、天地融合及星载计算的发展
The development of \ac{LEO} constellations and their integration with terrestrial mobile networks are extending global connectivity~\cite{ye2026wireless}.
% ~\cite{ye2026wireless, geng2026hierarchical}.
Alongside this networking progress, onboard computing is enabling satellites to process observations and execute \ac{AI} applications in orbit~\cite{geng2026hierarchical}.
% 介绍在轨AI处理能力，并落到应急观测中的筛选与下传
In emergency \ac{EO}, these capabilities support image screening, compression, and prioritization before transmission.
% 明确模型开始决定哪些信息对地面可见
Under limited and intermittent downlink opportunities, onboard models increasingly determine which observations are retained, delivered, or discarded, directly shaping the information availability to ground users.

% 2. 揭示漂移如何影响任务交付与恢复证据
% 运行条件偏离训练环境，导致关键观测被压低优先级、非关键观测占满链路两条路径
This control over data visibility makes model reliability a mission-level concern.
Heterogeneity in geography, season, and disaster exposure means that data encountered at deployment can fall outside the training distribution, a form of distribution shift~\cite{ekim2025distribution}.
A drifted filter may suppress critical observations or flood the downlink with highly ranked noncritical products, causing drift-induced evidence displacement.
% 下传数据也用于诊断和更新，因此错误筛选可能同时损失任务证据和恢复证据
The resulting losses affect both immediate disaster response and the evidence available for ground diagnosis and model adaptation.
Model-dependent selection can consequently weaken the evidence needed to identify and correct filtering errors~\cite{taori2023data}.
% 3. 概括已有进展，定位生命周期的连接缺口
% 概括已有工作在内容感知下传、星地模型适配和自主编排方面的进展
Existing studies support content-aware downlink, ground-assisted model adaptation, and autonomous task orchestration~\cite{herec2026sttorm,mateo2023orbit,gizzi2025onair,ye2026spacegrid}.
% 需要通过统一的证据流与权限机制，将这些能力组织为连续生命周期
However, their integration into a continuous lifecycle requires coordinated evidence exchange and explicit rules for initiating protection and restoring a model's data-disposition authority.

% 4. 确立全生命周期视角，引入Orbit-AWARE——贡献
% 长期有效性需要协同管理可靠性、证据和恢复权限
Sustainable onboard intelligence requires joint management of model reliability, evidence visibility, communication resources, and recovery authority.
% 介绍Orbit-AWARE的三平面架构
We therefore propose Orbit-AWARE, a lifecycle architecture whose cooperative Data, Cognitive, and Trust Planes coordinate continuous data processing, event-triggered agentic orchestration, and deterministic governance, respectively.
% 保护优先、事件触发编排与可信重新授权
The architecture protects observations before degradation is fully confirmed, organizes evidence exchange to support both immediate mission needs and subsequent recovery, and restores model authority after verification.
% 本文结构与贡献
% An illustrative emergency \ac{EO} case study examines this lifecycle under a common downlink budget, followed by future research directions toward self-recovering onboard intelligence.
The main contributions of this article are summarized as follows:
\begin{itemize}
    \item We characterize drift-induced evidence displacement through critical-observation suppression and priority flooding, linking mission delivery to model recoverability.
    We compare the lifecycle capabilities of representative approaches and derive management requirements from the identified gaps.
    \item We develop a safeguard-first agentic lifecycle that connects monitoring, protection, diagnosis, and model recovery with verified restoration of data-disposition authority.
    \item We present evidence-centric communication principles that coordinate what is retained and exchanged to support immediate mission needs, diagnostic assessment, and subsequent model adaptation.
    \item We provide an illustrative emergency \ac{EO} case study that evaluates evidence delivery, trusted model admission, and sustainable mission performance under a common communication budget.
\end{itemize}

\section{Onboard AI Data Filtering: Advances and Lifecycle Challenges}
\begin{figure*}
    \centering
    \includegraphics[width=0.95\textwidth]{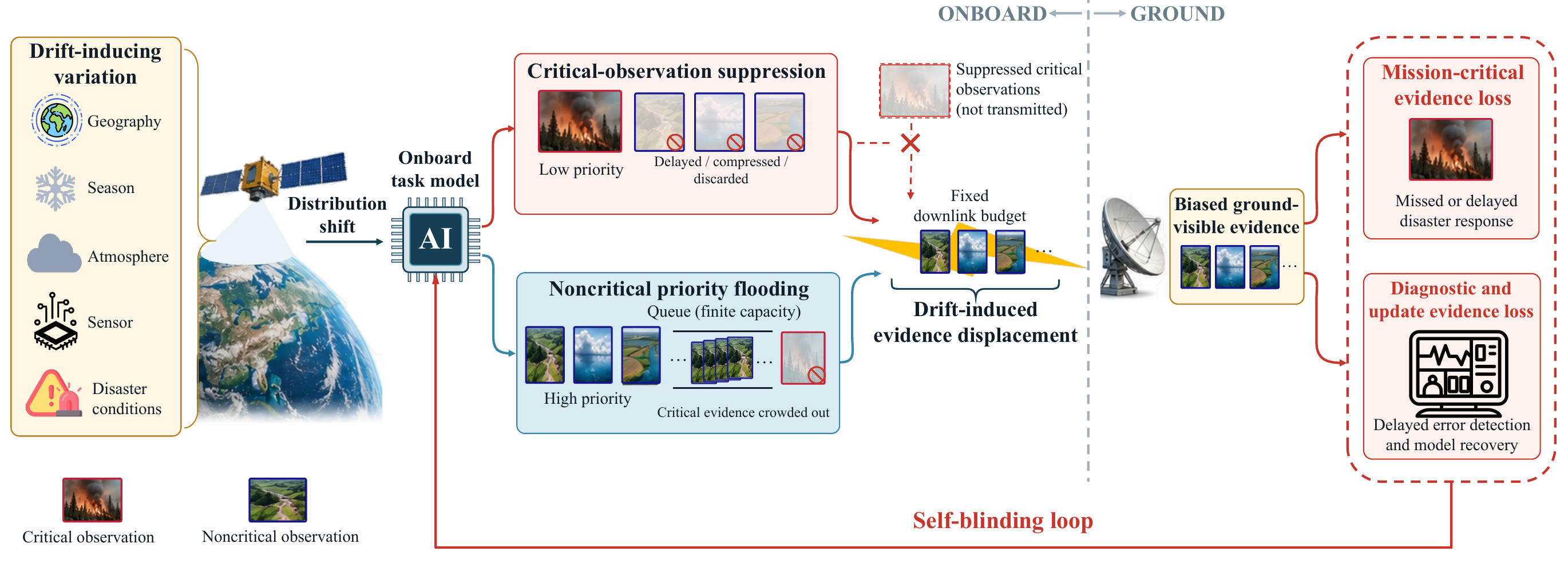}
    \caption{Drift-induced evidence displacement in bandwidth-limited emergency Earth observation.}
    \label{fig:evidence-displacement}
\end{figure*}
% AI获得数据处置权
% → 漂移被有限链路放大为证据置换
% → 点式方案无法保持可观察性并闭合恢复
% → 需要Data–Cognitive–Trust全生命周期架构
% 四个标题依次对应技术基础、失效机制、研究现状和架构需求
This section first reviews advances in \ac{AI}-enabled onboard data filtering,
then examines drift-induced evidence displacement and the limitations of existing solutions,
and finally derives the lifecycle requirements that motivate the proposed Orbit-AWARE architecture.

\subsection{AI-Enabled Onboard Data Filtering}
% 模型从推理器演变为数据筛选与采样节点
% 带宽矛盾与星上筛选；
% 数据处置权及三重角色；
% 应急观测的不可逆后果。
% 末句从“模型判断可能错误”转向“错误会改变未来可见数据”。
% Recent onboard \ac{AI} systems range from on-orbit validated cloud screening algorithms to more capable selective downlink schemes.
Recent onboard \ac{AI} systems support cloud screening, change-based prioritization, and selective downlink.
The $\Phi$-Sat-1 mission demonstrated that a compact \ac{DNN} can identify cloud-covered imagery before transmission~\cite{giuffrida2021varphi}. 
Other onboard change-detection pipelines have identified and encoded changed regions rather than repeatedly transmitting complete observations~\cite{bui2023board}. 
STTORM-CD advances this direction by fine-tuning a compact Siamese encoder with triplet loss to distinguish flood-related change from seasonal variation. It ranks image tiles for transmission under a constrained downlink budget while retaining compressed historical embeddings for onboard comparison~\cite{herec2026sttorm}. 
Recent satellite--ground vision-language systems use confidence-guided inference offloading and task-aware regional compression to reduce transmission demand~\cite{zhang2025satellite}.
% ~\cite{zhang2025satellite,park2026summarize}.
These developments mark a shift from raw-data relay toward content-aware data handling.

% Thus, onboard filtering encompasses a continuum of decisions that governs data retention, feature representation, transmission scheduling, and bandwidth allocation. 
% 压缩字数
Onboard filtering determines data retention, representation, and transmission under intermittent links and onboard resource constraints.
% Onboard filtering determines which observations are retained, how they are represented, and when they are transmitted.
% % under the available bandwidth,
% It supports ground analysis by preserving and prioritizing useful information under intermittent links and onboard resource constraints.

Once the model outputs control retention and transmission, the task-oriented model performs three coupled functions: scene interpretation, communication priority ranking, and implicit sampling for subsequent model diagnosis and training. 
% This coupling is critical for emergency observations of rare and time-sensitive disaster events, which are difficult to reacquire. 
This coupling is particularly consequential in disaster response, where observations of rapidly evolving conditions may be impossible to reacquire.
Accordingly, unreliable onboard decisions can impair both timely delivery of critical observations and the evidence base for ground analysis and model maintenance.
Such issues become persistent lifecycle risks when on-orbit operational environments deviate from the model's training domain.

\subsection{Drift-Induced Evidence Displacement}
% Related works
% 区分shift、reliability question与degradation；
% 解释suppression和flooding；
% 解释选择偏差与self-blinding
The shift from inference to data disposition changes how model drift should be interpreted.
Geographic, seasonal, atmospheric, sensor, and disaster-specific variations can move observations away from the training distribution~\cite{ekim2025distribution}. 
% as recent \ac{EO} studies have documented~\cite{ekim2025distribution}.
Yet such a shift does not by itself prove task degradation.
We distinguish three operational states.
Distribution shift denotes a change in incoming data.
Questioned reliability arises when that change coincides with inconsistent model behavior.
Confirmed degradation requires labeled or external evidence of a loss in task performance.
% This distinction avoids treating every detected anomaly as a model failure.
These operational states guide the transition from routine monitoring to precautionary protection and evidence-based recovery.

Under a limited downlink budget, ranking errors distort resource allocation through the two paths illustrated in Fig.~\ref{fig:evidence-displacement}.
In the \textit{suppression} path, critical observations receive low scores and are compressed, delayed, or discarded.
In the \textit{flooding} path, noncritical observations receive excessive priority, occupy scarce queue positions and transmission capacity, and crowd out critical observations.
We refer to their common outcome as drift-induced evidence displacement: drift distorts the allocation of a fixed communication budget and thereby changes which evidence reaches the ground.
This effect may remain hidden when evaluation considers aggregate predictive accuracy but not the composition of transmitted data.

% The loss extends beyond the immediate disaster product.
Downlinked observations support ground verification, failure diagnosis, and subsequent model updates.
Model-dependent data selection can amplify existing dataset biases~\cite{taori2023data},
because observations excluded by the filter may remain unavailable for verification or learning.
By the next ground contact, discarded observations may be unavailable, and the original event conditions may no longer be observable.
When the displaced observations would also have exposed or corrected the failing behavior, the system enters a self-blinding loop,
where unreliable filtering removes part of the evidence required to recognize and recover from that unreliability.
The challenge is therefore to detect distribution shift, preserve counter-evidence---observations that challenge the current ranking or are retained outside its selection process---and maintain a path to recovery.

\subsection{Existing Solutions and Limitations}
% 引导段说明各研究流分别贡献了闭环组件；
% Table I进行范式级比较；
% 综合段强调缺口来自组件之间缺少时序、信息和权限连接，而不是缺少某个单独算法
\begin{table*}[!tb]
\centering
\caption{Lifecycle capability coverage of representative research directions and Orbit-AWARE.}
\label{tab:lifecycle-coverage}
\begingroup
\small
\setlength{\tabcolsep}{3pt}
\renewcommand{\arraystretch}{1.3}
\setlength{\extrarowheight}{2pt}
% Keep the column shading continuous across rows and rule boundaries.
\setlength{\aboverulesep}{0pt}
\setlength{\belowrulesep}{0pt}
\def\OAcolwidth{\dimexpr(\textwidth-.34\textwidth-14\tabcolsep)/7\relax}
\def\OAfull{\ensuremath{\checkmark}}
\def\OApartial{\ensuremath{\triangle}}
\def\OAoutside{\ensuremath{\times}}
\begin{tabular}{@{}
>{\raggedright\arraybackslash}m{.34\textwidth}
*{6}{>{\centering\arraybackslash}m{\OAcolwidth}}
>{\columncolor{gray!10}[\tabcolsep][0pt]\centering\arraybackslash}m{\OAcolwidth}
@{}}
\toprule
\textbf{Lifecycle capability} &
\shortstack{\textbf{Onboard}\\\textbf{filtering}} &
\shortstack{\textbf{Semantic}\\\textbf{downlink}} &
\shortstack{\textbf{OOD}\\\textbf{detection}} &
\shortstack{\textbf{Ground-}\\\textbf{assisted}\\\textbf{retraining}} &
\shortstack{\textbf{Federated}\\\textbf{adaptation}} &
\shortstack{\textbf{Cognitive/}\\\textbf{agentic}\\\textbf{control}} &
\shortstack{\textbf{Orbit-}\\\textbf{AWARE}\\{\scriptsize\textbf{(This work)}}}\rule[-6pt]{0pt}{40pt} \\
\midrule
\textbf{1. Sense: Reliability monitoring}\newline
\textit{Tracks input changes and filtering behavior.} &
\OApartial & \OAoutside & \OAfull & \OApartial & \OAoutside & \OApartial & \OAfull \\
\textbf{2. Awaken: Event-triggered orchestration}\newline
\textit{Activates control or cognitive tools after an event.} &
\OAoutside & \OAoutside & \OApartial & \OAoutside & \OAoutside & \OAfull & \OAfull \\
\textbf{3. Safeguard: Counter-evidence protection}\newline
\textit{Preserves observations outside score-based selection.} &
\OAoutside & \OAoutside & \OAoutside & \OAoutside & \OAoutside & \OApartial & \OAfull \\
\textbf{4. Diagnose: Evidence-guided assessment}\newline
\textit{Examines causes and task implications of change.} &
\OAoutside & \OAoutside & \OApartial & \OApartial & \OAoutside & \OApartial & \OAfull \\
\textbf{5. Curate and Exchange: Evidence selection}\newline
\textit{Selects task or learning information for transmission.} &
\OAfull & \OAfull & \OAoutside & \OApartial & \OApartial & \OApartial & \OAfull \\
\textbf{6. Recover: Model adaptation}\newline
\textit{Updates the model using accumulated learning evidence.} &
\OAoutside & \OAoutside & \OAoutside & \OAfull & \OAfull & \OApartial & \OAfull \\
\textbf{7. Assure and Resume: Governed reactivation}\newline
\textit{Validates candidates and controls model readmission.} &
\OAoutside & \OAoutside & \OAoutside & \OApartial & \OApartial & \OApartial & \OAfull \\
\midrule
\textbf{Representative references} &
\cite{giuffrida2021varphi,bui2023board,herec2026sttorm} &
\cite{zhang2025satellite} &
\cite{ekim2025distribution} &
\cite{mateo2023orbit} &
\cite{zhang2024leader} &
\cite{gizzi2025onair,li2024agent} &
\textbf{This work} \\
\bottomrule
\end{tabular}
\par\smallskip
\begin{minipage}{\textwidth}
\footnotesize
\OAfull: explicitly addressed; \OApartial: component-level or application-dependent coverage; \OAoutside: outside the primary scope of the cited representatives.
Entries summarize the cited works, not every method within a research direction.
The Orbit-AWARE column indicates the lifecycle capabilities specified by the architecture.
\end{minipage}
\endgroup
\end{table*}
Existing studies provide several components needed to address the above challenge, but they intervene at different points in the operational
sequence.
Onboard filtering and semantic downlink reduce transmission volume, whereas \ac{OOD} detection indicates when incoming observations differ from a reference distribution.
Neither capability by itself ensures that observations conflicting with unreliable filter outputs are retained for diagnosis.
Recovery-oriented studies begin later in the sequence.
The WorldFloods payload demonstrated that images can be downlinked for ground labeling and retraining before updated weights are uploaded to the spacecraft~\cite{mateo2023orbit},
while satellite \ac{FL} enables models to be updated from distributed onboard data~\cite{zhang2024leader}.
These approaches support adaptation from available learning evidence, 
but evidence preservation is not integrated into their filtering policies.

Autonomy frameworks contribute a complementary capability.
OnAIR offers a modular cognitive architecture for integrating onboard \ac{AI} applications and mission tools~\cite{gizzi2025onair}, 
and Agent-as-a-Service organizes planning, execution, resource management, and long-term evolution as an agent-managed service lifecycle~\cite{li2024agent}.
Such orchestration can coordinate heterogeneous tools, but it does not on its own define when a drifting task model should lose data-disposition authority or what evidence is required before that authority is restored.
% Table~\ref{tab:lifecycle-coverage} summarizes these complementary research directions at the lifecycle level rather than comparing individual algorithms.
Table~\ref{tab:lifecycle-coverage} maps these research directions to lifecycle stages and identifies integration gaps.

\subsection{Requirements for Lifecycle Management}
% 第一段提出R1–R4；
% 第二段将要求映射到Data、Cognitive、Trust三平面，并自然引出Section III
In summary, the lifecycle gaps identified above lead to four requirements for an onboard model that controls data retention and transmission.
First, \textit{persistent observability} is needed to track changes in both incoming observations and filtering behavior.
Because the task model determines which data remain visible, its reliability cannot be assessed solely from its own scores or selected outputs.
Independent distribution cues, protected samples, and operational records must remain available when its decisions become questionable.
Second, the system must \textit{safeguard before certainty}.
A shift alarm does not establish task degradation, but waiting for ground confirmation may allow rare and diagnostic observations to be discarded.
Reversible measures, including exploratory retention and temporary protection of uncertain observations, should therefore limit evidence loss while reliability is being established.
Third, \textit{evidence-centric coordination} must balance immediate mission value and future diagnostic value under a fixed communication budget.
Evidence exchanged among the target satellite, its peers, and the ground segment should support event reporting, degradation assessment, and model correction beyond selection based on model confidence.
Fourth, recovery must be \textit{governed}.
Model diagnosis, parameter adaptation, and restoration of data-disposition authority are distinct decisions, since an updated model may improve on available samples without being reliable across subsequent observations.
Reactivation therefore requires validation evidence, staged operation, and a rollback path.

% Taken together, these four requirements indicate that an onboard \ac{AI} lifecycle management system should comprise at least three functional planes.
% Persistent observability maps primarily to the \textbf{Data Plane}, while evidence protection and coordination span the \textbf{Cognitive Plane}.
% Recovery authority is assigned to the \textbf{Trust Plane}.
These requirements call for an onboard \ac{AI} \ac{LCM} architecture that separates high-rate data handling, event-triggered cognitive orchestration, and deterministic lifecycle governance while allowing evidence to flow across them,
% 3GPP LCM
extending terrestrial \ac{AI} \ac{LCM} concepts toward resource‑constrained on‑orbit scenarios~\cite{3gpp.28.105}.

\section{Orbit-AWARE Architecture}
\label{sec3}
% \begin{figure*}[!t]
%     \centering
%     \includegraphics[width=0.98\textwidth]{In-orbit-WCM-1031/WCM1101/pdf/Fig2-temp.png}
%     \caption{Orbit-AWARE architecture for sustained in-orbit intelligence in emergency EO.
%     }
%     \label{fig:orbit-aware-architecture}
% \end{figure*}
% 全文核心，回答 Orbit-AWARE 如何管理在轨 AI 的完整生命周期
This section first presents the design principles of Orbit-AWARE, 
then describes its three cooperative planes,
and finally explains the safeguard-first agentic lifecycle.

\subsection{Design Principles}
% safeguard before certainty
% event-triggered cognition
\begin{figure*}[!t]
    \centering
    \includegraphics[width=0.98\textwidth]{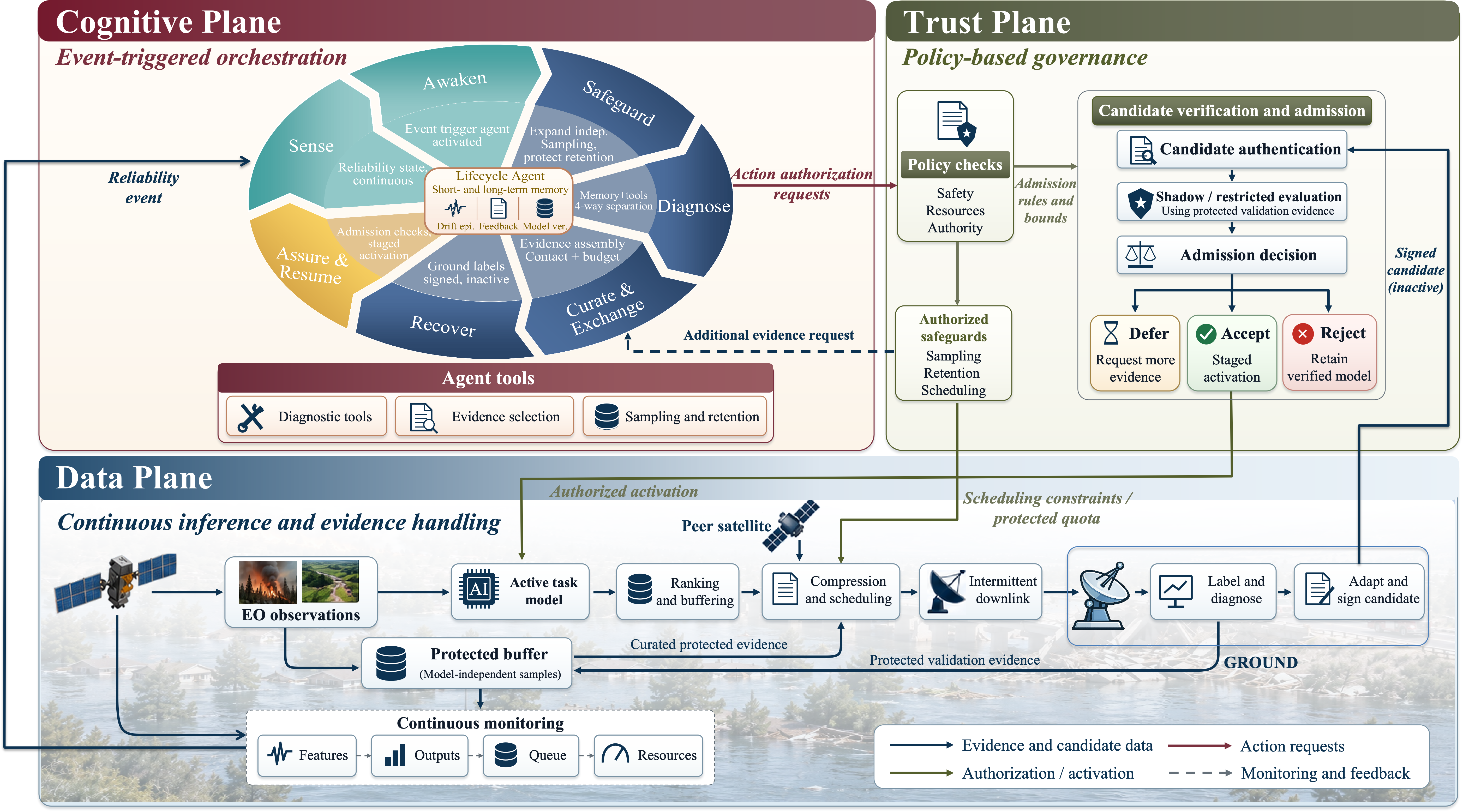}
    % \caption{Orbit-AWARE architecture for sustainable in-orbit intelligence.
    % }
    \caption{Orbit-AWARE architecture and safeguard-first lifecycle. Continuous monitoring in the Data Plane supports event-triggered orchestration in the Cognitive Plane. The Trust Plane authorizes protective actions and model admission, while the ground segment supplies labels and signed candidate updates.}
    \label{fig:orbit-aware-architecture}
\end{figure*}
Orbit-AWARE separates continuous task execution from episodic lifecycle intervention.
During nominal operation, the onboard filter processes observation streams while lightweight monitors track changes in data representations, model outputs, and queue composition.
When filtering reliability is questioned, the cognitive agent is awakened to examine the event while predefined reversible safeguards preserve evidence pending diagnosis.
Computationally intensive reasoning therefore remains outside the routine inference path.
% When filtering reliability is questioned, reversible safeguards preserve evidence before task degradation is fully confirmed.
% measures such as protected buffering, increased independent sampling, and temporary restrictions on irreversible data disposal 
% This safeguard-first response preserves observations that may otherwise disappear while the cause and severity of the change are still being established.
% Computationally intensive reasoning is reserved for such events, 
% keeping routine inference separate from episodic lifecycle intervention.
% Orbit-AWARE therefore follows two principles: 
% evidence is safeguarded before diagnostic certainty is reached, 
% and cognitive orchestration is invoked only when the reliability state warrants intervention.
% The cognitive agent is then invoked to examine the event, keeping computationally intensive reasoning outside the routine inference path.

% Once awakened, the cognitive agent operates within an evidence- and policy-bounded authority model.
% Evidence-grounded agency requires every proposed action to be traceable to observable system states and verified tool outputs.
% by structured monitoring states, historical feedback, model lineage, and verified tool outputs.
Once awakened, the cognitive agent proposes actions based on observable system states and verified tool outputs.
Each proposal is recorded with its supporting evidence and checked against predefined policies.
Deterministic policies retain authority over resource commitments, irreversible data disposition, and model activation.
% The resulting proposals may adjust retention, sampling, and evidence exchange, while actions affecting irreversible data disposal, resource limits, or model activation are independently checked against deterministic policies.
The ground segment remains responsible for labeling, model adaptation, and candidate signing.
Onboard data-disposition authority is restored after the candidate satisfies the validation and admission requirements.
% Trust-governed autonomy therefore varies with the available evidence: routine filtering proceeds locally, uncertain conditions invoke protection and diagnosis, and model reactivation requires explicit authorization.
% These complementary responsibilities are assigned to the Data, Cognitive, and Trust Planes introduced in the following subsection.

\subsection{Three Cooperative Planes}
% 结合Fig.1展开应急 EO 场景、星—地—星拓扑、Data–Cognitive–Trust 三平面、Agentic Lifecycle 和可信恢复路径
% 重点解释三平面如何共同维持：
% - 模型可观测性；
% - 关键数据可见性；
% - 恢复证据连续性；
% - 模型版本与数据处置权的可信切换。
% The preceding design principles are realized through the Data, Cognitive, and Trust Planes illustrated in Fig.~\ref{fig:orbit-aware-architecture}.
% In the emergency \ac{EO} workflow, incoming observations are continuously processed in the Data Plane, where task products compete for limited storage and downlink opportunities.
% When filtering reliability is questioned, monitoring states, protected samples, and resource information are passed to the Cognitive Plane, where the event-triggered agent uses memory and diagnostic tools to formulate evidence-management and recovery actions.
% % These actions, together with candidate models returned by the ground segment, are evaluated by the Trust Plane before they can alter irreversible data disposition or active model states.
% The Trust Plane checks the proposed actions and candidate models before authorizing irreversible data-disposition decisions or changes to the active model.
% The separation allows high-rate task processing, adaptive cognitive reasoning, and deterministic lifecycle governance to operate at their appropriate timescales while remaining connected through evidence exchange.
The three planes in Fig.~\ref{fig:orbit-aware-architecture} separate continuous data processing, event-triggered orchestration, and lifecycle authorization.
Monitoring states and protected observations inform the cognitive agent, while action requests and admission outcomes connect its decisions to policy-governed execution.

The Data Plane provides the continuous execution layer of Orbit-AWARE.
It hosts the task model and performs observation inference, priority ranking, buffering, compression, and transmission scheduling at the acquisition rate.
Lightweight monitors track representation statistics, output behavior, queue composition, and resource availability without interrupting the normal filtering path.
% To prevent the task model from defining its own observation boundary, a minimum model-independent sampling channel and a protected evidence buffer are maintained outside the score-based ranking process.
% Consequently, observations that contradict the current model can remain available even when they would not be selected for routine downlink.
A baseline stream of model-independent samples is retained in a protected buffer outside the score-based selection process.
Together with monitoring summaries and resource states,
these otherwise-excluded observations support event-level diagnosis and policy adjustment in the Cognitive Plane.

On this evidence base, the Cognitive Plane performs event-triggered diagnosis and evidence orchestration.
Once awakened, the agent combines current monitoring states with short- and long-term memory, including previous drift episodes, delayed ground feedback, and model-version histories.
% Diagnostic tools are invoked to distinguish benign domain variation from sensor anomalies, ranking failures, and persistent task degradation,
% and then the resulting diagnosis is translated into bounded actions, 
% such as increasing independent sampling, protecting selected observations, changing compression levels, or requesting complementary evidence from the ground or peer satellites.
Diagnostic tools help distinguish benign domain variation from sensor anomalies, ranking failures, and persistent task degradation.
The agent uses these findings to propose adjustments to sampling, retention, compression, or to request complementary evidence from the ground or peer satellites.
% As new feedback becomes available, these actions can be revised across successive contact windows rather than being fixed at the initial alarm.
% Each proposed action is recorded with its supporting evidence and submitted to the Trust Plane for authorization.
Actions are revised as feedback arrives across contact windows and submitted with supporting evidence to the Trust Plane for authorization.

% The Trust Plane evaluates the proposed actions against deterministic safety, resource, and authority policies.
% % Adjustments to sampling, buffering, and transmission are admitted only when their resource demands and validity periods remain within predefined bounds, while irreversible deletion and model activation require stronger authorization.
% Pre-authorized adjustments to sampling, buffering, and transmission can be executed within predefined resource bounds and validity periods.
% Irreversible deletion and model activation require explicit authorization.
% A candidate model returned by the ground segment is authenticated and evaluated in a shadow or restricted operating mode using protected validation evidence.
% % Depending on the amount and consistency of that evidence, reactivation can be deferred, accepted in stages, or rejected through rollback to the last verified model.
% Admission is deferred when evidence is insufficient, and candidates that satisfy the requirements proceed to staged activation.
% Candidates that fail the checks are rejected, while the last verified model remains available for rollback after activation.
% Model versions, policy decisions, and execution outcomes are retained as auditable records and returned to the Cognitive Plane as lifecycle feedback.
% The temporal interaction of the three planes is organized by the safeguard-first agentic lifecycle described next.
% 突出Trust Plane的核心价值——管理动作执行与模型处置权限。Defer、Reject、Accept的完整流程集中保留在III-C
The Trust Plane governs which actions may be executed and which model may control data disposition.
Pre-authorized sampling, retention, and scheduling adjustments operate within predefined resource bounds and validity periods.
Candidate authentication and protected validation govern model admission, while version records preserve a rollback path.
Decisions and execution outcomes are returned to the Cognitive Plane as auditable lifecycle feedback.

\subsection{Safeguard-First Agentic Lifecycle}
Orbit-AWARE implements a safeguard-first, seven-stage lifecycle
that combines continuous sensing with a six-stage event-triggered agentic loop,
as illustrated in Fig.~\ref{fig:orbit-aware-architecture}.
% that combines continuous sensing with a six-stage event-triggered agentic loop to protect mission and recovery evidence before model degradation is fully diagnosed.
During nominal operation, \textit{Sense} combines feature statistics, model outputs, queue behavior, and independently sampled observations into a compact reliability state.
% in the Data Plane, integrating changes in feature distributions, confidence behavior, model disagreement, queue composition, and counter-evidence obtained through model-independent exploration into a compact reliability state.
% This state indicates whether the current importance ranking remains a dependable basis for data disposition.
When this state calls the current ranking into question, 
\textit{Awaken} activates the event-triggered agent, 
and \textit{Safeguard} immediately expands model-independent sampling and protected retention.
% When multiple signals challenge that reliability, \textit{Awaken} activates the event-triggered agent and retrieves the relevant model history, prior events, and available tool interfaces.
% Before a complete diagnosis is reached, \textit{Safeguard} expands independent sampling, reserves protected buffer capacity, and suspends irreversible deletion for novel or conflicting observations.
This transition preserves novel or conflicting observations before diagnosis is complete,
limiting critical-evidence suppression and priority flooding during the response interval.
% Placing evidence preservation ahead of diagnostic certainty limits both critical-evidence suppression and priority flooding during the interval in which the cause and severity of drift remain under examination.

With protected evidence in place, \textit{Diagnose} uses lifecycle memory and analytical tools to distinguish benign domain variation from emerging disaster patterns, sensor anomalies, and filter degradation.
% Once the safeguard is in place, the lifecycle shifts from precautionary protection to evidence-guided recovery.
% During \textit{Diagnose}, the agent uses model history, event memory, and analytical tools to distinguish benign domain variation from emerging disaster patterns, sensor anomalies, and genuine filter degradation.
The identified diagnostic and knowledge gaps then drive \textit{Curate and Exchange}, which assembles task- and recovery-relevant evidence under the available contact opportunities and communication budget.
% The remaining diagnostic and knowledge gaps then guide \textit{Curate and Exchange}, where observations, regions of interest, feature summaries, and peer-satellite knowledge are ranked according to their immediate mission value, diagnostic relevance, recovery value, and communication cost.
During \textit{Recover}, the ground segment combines the delivered evidence with labels and training resources to generate a signed candidate model or lightweight adaptation package.
% The resulting evidence packages are delivered whenever ground contacts or \ac{ISLs} become available, allowing scarce transmission opportunities to serve both current disaster response and subsequent model repair.
% During \textit{Recover}, the ground segment combines the received evidence with labels and training resources to produce a signed candidate model or lightweight adaptation package.
This ground-first path maintains time-critical data disposition onboard while the returned candidate remains inactive pending Trust Plane admission.
% This ground-first recovery path separates time-critical onboard data disposition from compute-intensive model adaptation, while the returned candidate remains inactive until it is admitted by the Trust Plane.

The inactive candidate then enters \textit{Assure and Resume}, where the Trust Plane verifies its authenticity and evaluates its performance in shadow or restricted operation using protected validation evidence.
Admission is governed by established performance bounds and mission-specific rules.
An admitted candidate proceeds to staged activation before full data-disposition authority is restored.
When the evidence remains insufficient, admission is deferred and additional observations are requested, 
whereas a candidate that violates the required bounds is rejected and the last verified model remains active.
These deterministic decisions separate model reactivation from the agent's diagnostic and planning functions.
Admission outcomes and model states are recorded in lifecycle memory and returned to the Data Plane, 
updating the context for continuous \textit{Sense} and subsequent interventions.
% The inactive candidate enters \textit{Assure and Resume}, where the Trust Plane evaluates it against protected validation evidence, established performance bounds, and mission-specific admission rules.
% A candidate that satisfies the admission criteria is first introduced through shadow evaluation or staged activation before its data-disposition authority is restored.
% If the available evidence is insufficient, deployment is deferred while additional observations are requested, whereas a candidate that violates the required bounds is rejected and the verified model remains in service.
% These deterministic decisions keep model reactivation separate from the agent’s diagnostic and planning functions, ensuring that cognitive autonomy does not bypass lifecycle governance.
% The admission outcome, supporting evidence, and updated model state are then recorded in lifecycle memory and fed back to the Data Plane.
% Orbit-AWARE consequently returns to continuous \textit{Sense} with a revised reliability baseline, closing the loop between drift awareness, evidence preservation, recovery, and trusted reactivation.

\section{Evidence-Centric Communications}
% 通信核心：何时改变通信策略、选择哪些证据、如何利用协同反馈完成恢复
% A：可靠性事件决定何时介入 → B：诊断需求决定交换什么 → C：协同反馈决定下一步如何推进
This section first explains reliability-aware triggering, 
then discusses evidence selection over limited links, 
and finally describes space--ground--peer coordination across the Orbit-AWARE lifecycle.

\subsection{Reliability-Aware Triggering}
% 连续监测产生变化信号
% → 多源证据使筛选可靠性受到质疑
% → 触发保护与认知介入
% → 根据后续证据调整介入程度

% 说明Sense如何形成可供决策的可靠性证据
Reliability-aware triggering connects \textit{Sense}, \textit{Awaken}, and \textit{Safeguard} by determining when routine filtering requires intervention.
Changes in input or feature distributions provide an initial indication that operating conditions have shifted, 
while output behavior and queue composition reveal how that shift may affect transmission priorities.
For example, a growing concentration of high-scoring products may reflect either intensified disaster activity or excessive priority assigned to noncritical observations.
Model-independent samples and external feedback, when available, help distinguish these possibilities by exposing inconsistencies that score-based monitoring may overlook.
% Persistent agreement among these complementary signals establishes a reliability event that activates cognitive assessment and precautionary protection before task degradation is fully confirmed.
When complementary signals consistently indicate that the ranking may be unreliable, an alert triggers cognitive assessment and precautionary protection before task degradation is fully confirmed.

% 说明Awaken与Safeguard如何改变通信和保留策略
Upon this reliability event, \textit{Awaken} initiates agent-led assessment while \textit{Safeguard} applies predefined, reversible protection within the available resource budget.
Independent sampling is increased, and selected novel or conflicting observations are retained for subsequent examination.
This protected retention preserves future selection opportunities, while actual transmission remains subject to the link budget and contact schedule.
The agent then uses the event context to formulate ground or peer verification requests and identify the additional evidence needed for diagnosis.
As feedback accumulates, protection can be extended, refined, or released according to the evolving reliability state.
% The resulting intervention therefore progresses from immediate evidence preservation to targeted exchange, making the choice of what to transmit the next communication decision.
These verification requests guide evidence selection during \textit{Curate and Exchange}.

\subsection{Recovery Evidence over Limited Links}
% Diagnose识别尚未解决的问题
% → 将问题转化为证据需求
% → 选择证据内容与表示形式
% → 在任务交付和恢复需求之间分配链路

% 说明Diagnose的输出如何决定Curate and Exchange的对象
% The choice of what to transmit follows from the questions left unresolved by \textit{Diagnose}, linking its findings to \textit{Curate and Exchange}.
During \textit{Curate and Exchange}, evidence is selected to answer the unresolved questions identified in \textit{Diagnose}.
Health statistics and score-distribution summaries can help the ground segment assess whether an anomaly persists across observation windows, 
whereas representative images and conflicting samples support examination of unfamiliar disaster patterns or ranking errors.
% When model adaptation is required, the evidence package must also preserve the visual detail and acquisition context needed for labeling and training.
% The representation is therefore selected according to its intended use: 
Compact summaries support initial assessment, while image tiles or regions of interest preserve the detail needed for visual verification, labeling, and training.
Acquisition metadata preserve context for ground interpretation.

% 说明有限链路中的证据取舍和动态分配
Once these evidence needs are identified, the available transmission capacity is allocated between time-sensitive mission products and observations that can support diagnosis or recovery.
Candidate observations are assessed for task relevance, novelty, model disagreement, and coverage of unfamiliar conditions.
Their expected contribution is weighed against the transmission cost.
The same observation may support both disaster response and model recovery,
while complementary samples can expose ranking errors missed by selecting only the highest-scoring products.
A share of the transmission budget is therefore reserved for model-independent exploration, preserving access to evidence outside the current ranking.
The agent uses lifecycle memory, tool outputs, and ground feedback to revise sample selection and representation as diagnostic needs and contact opportunities change.
% the agent revises the evidence mix and representation as diagnostic needs, contact opportunities, and ground feedback change.
% Each exchange consequently addresses the current information gap and provides a basis for deciding what should be requested or delivered next.

\subsection{Space–Ground–Peer Coordination}
% 证据交换产生核验反馈
% → 反馈修订后续取样需求
% → 累计证据支持模型恢复
% → 准入结果返回下一轮感知与选择

% 解释断续连接下的星地请求—反馈过程
Space--ground coordination links \textit{Curate and Exchange} to \textit{Recover} through successive rounds of evidence delivery and feedback.
An initial exchange provides a compact account of the reliability event together with representative observations.
The ground segment examines this evidence and returns verified labels, diagnostic findings, or requests for samples from conditions that remain poorly represented.
These responses are incorporated into the agent's memory and used to refine subsequent sampling and transmission decisions across contact windows.
Acquisition time, location, model version, and selection history preserve observation provenance for joint ground analysis across contacts.
% Through this process, the accumulated evidence progressively supports targeted model adaptation while ground feedback guides the use of subsequent downlink opportunities.

% 解释邻星补充、恢复返回及闭环连续性
Peer satellites can contribute distribution summaries, complementary observations, or reports of prediction disagreements from related regions and operating conditions.
Their relevance is assessed using observation time, sensing conditions, and model-version information before they are incorporated into the evidence package.
Once adaptation is completed, the ground segment returns a signed candidate package with its version and validation records, connecting \textit{Recover} to \textit{Assure and Resume}.
% If evidence is insufficient, further observations are requested.
% If the candidate is rejected, the verified model remains active.
% Following acceptance, the new model version and admission record become part of the context used by \textit{Sense} and subsequent evidence selection.
Admission outcomes and requests for additional evidence are exchanged as lifecycle feedback, informing subsequent sensing and evidence selection.
% These outcomes are recorded in lifecycle memory and provide context for the next \textit{Sense} cycle and subsequent evidence selection.
% The following case study examines how this communication-supported lifecycle affects evidence delivery, trusted recovery, and sustained mission utility under a common downlink budget.

\section{Case Study: Sustainable Onboard Intelligence under Drift}
% 数据集与漂移构造、对比方法、公平性约束、指标公式、结果图表、消融、结果解释和适用边界
This section first introduces the experimental setup, 
then evaluates evidence delivery under drift, 
and finally examines trusted recovery and sustainable mission utility under a common downlink budget.

\subsection{Experimental Setup}
We consider a bandwidth-limited emergency \ac{EO} task built on the xBD disaster-damage dataset~\cite{gupta2019creating}.
% Although xBD provides paired pre- and post-disaster satellite imagery with building-level damage annotations, 
Post-disaster image tiles compete for the available downlink capacity.
A tile is regarded as a critical observation if it contains at least one building annotated as having major damage or being destroyed.
A ResNet-18 trained on source-disaster events serves as the initial onboard filter, while the \textit{Socal-fire} event is reserved as an unseen deployment domain.
The resulting importance score determines the transmission order of incoming products, making the model both a task classifier and a gatekeeper of ground-visible evidence.

To examine how this gatekeeper behaves as ranking reliability deteriorates, we organize 577 real \textit{Socal-fire} products into an 18-window controlled deployment replay.
The products are first partitioned into critical and noncritical groups using tile-level labels derived from the original building damage annotations.
Noncritical products are ordered by increasing frozen-model score, whereas critical products are ordered by decreasing score.
This ordering produces coupled priority flooding and critical-evidence suppression while keeping the class prevalence approximately stable across windows.
The original images, damage annotations, model scores, and feature embeddings remain unchanged, and each product appears exactly once.
Ground-truth labels are used to configure and evaluate the offline replay but remain unavailable to all deployable online policies.
% By controlling class prevalence and preserving the original model outputs, the replay isolates how deteriorating ranking reliability propagates to evidence delivery and subsequent model recovery.

Under this common replay, we compare Orbit-AWARE with four representative alternatives.
The Static Filter always follows the frozen-model ranking, whereas Fixed Exploration reserves a fixed share of each downlink window for model-independent sampling.
The Rule-Based Drift Loop activates a predefined uncertainty-sampling response after its monitoring criterion is met and a fixed response delay has elapsed.
% Orbit-AWARE instead responds at the reliability-alert stage and uses delayed labels from its own previous downlinks to revise evidence selection across windows.
A mission-oriented Oracle accesses current labels and serves as an idealized label-informed reference for critical-observation delivery under the same communication budget.
All methods process the same data stream with the same initial filter and a 10\% token-bucket budget, which permits 58 products, or 8.73~MB, over the deployment period.
All adaptive methods use the same ground-side adapter, and no deployable method accesses current labels.
% The comparison therefore evaluates how the lifecycle policies differ in protection timing and evidence selection.

\subsection{Evidence Delivery under Drift}
\begin{figure*}[!t]
    \centering
    \includegraphics[width=0.99\textwidth]{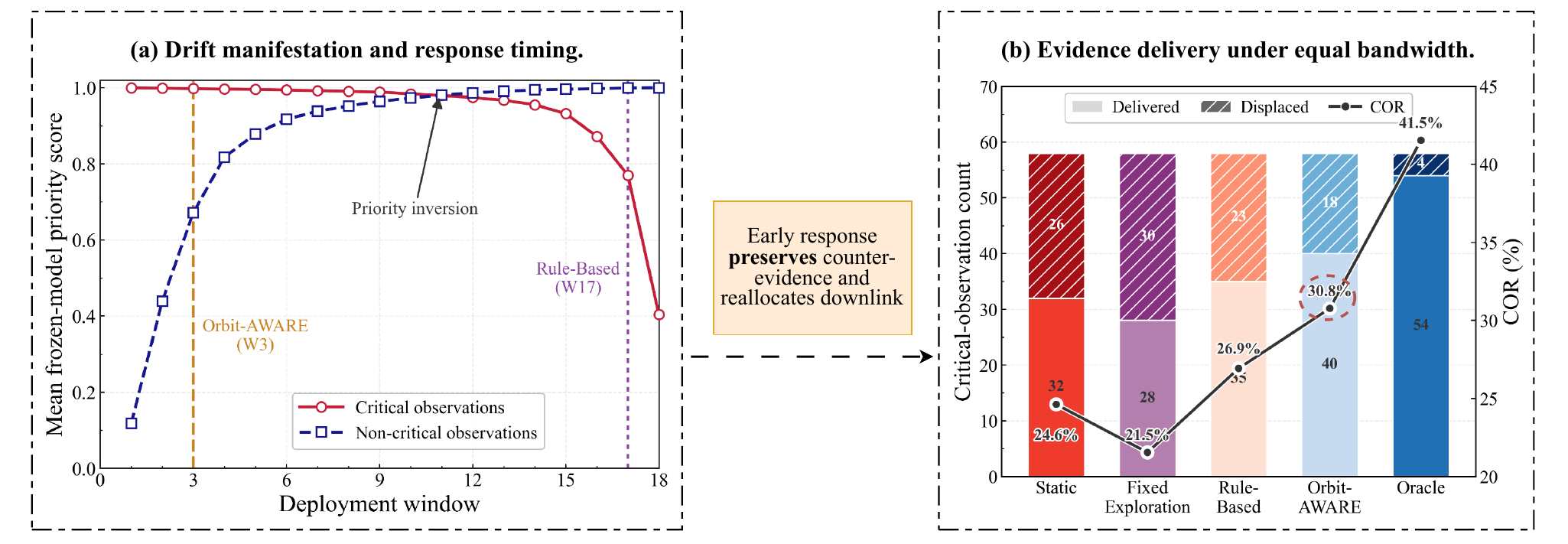}
    % \caption{Illustration of drift manifestation and evidence delivery in the xBD-based deployment replay.
    % }
    % \caption{Ranking reversal, safeguard activation, and critical-observation delivery in the xBD-based deployment replay.
    % }
    \caption{Ranking degradation and critical-evidence delivery in the controlled xBD replay: (a) class-wise mean frozen-model scores and safeguard activation windows; (b) delivered critical observations, the gap to the 58-product budget ceiling, and critical observation recall.}
    \label{fig:drift-delivery}
\end{figure*}
% The evolution of the frozen model's ranking behavior is illustrated in Fig.~\ref{fig:drift-delivery}(a).
% % 缩写，删除一个公式
% For each deployment window, the mean priority score is obtained by averaging
% the frozen-model scores separately over critical and noncritical
% observations.
% % For class $c$ in deployment window $w$, the mean priority score is defined as
% % \begin{equation}
% %     \bar{s}_{c,w}
% %     =
% %     \frac{1}{N_{c,w}}
% %     \sum_{i \in \mathcal{D}_{c,w}} s_i^{(0)},
% % \end{equation}
% % where $\mathcal{D}_{c,w}$ denotes the set of observations belonging to class $c$, $N_{c,w}=|\mathcal{D}_{c,w}|$, and $s_i^{(0)}$ is the priority score assigned by the frozen model.
% A progressive reversal between critical and noncritical observations 
% % of critical observations and noncritical ones 
% is observed over the deployment period.
The class-wise mean frozen-model scores in each deployment window are illustrated in Fig.~\ref{fig:drift-delivery}(a), revealing a progressive ranking reversal.
The mean score of critical observations decreases from $0.9995$ in W1 to $0.4042$ in W18, 
whereas that of noncritical observations increases from $0.1175$ to $0.9998$.
The two mean-score trajectories cross between W10 and W11, 
after which noncritical observations receive higher scores than critical observations.
% The Orbit-AWARE safeguard is activated at W3, before this priority inversion becomes visible in the task scores, while the Rule-Based Drift Loop is activated at W17.
% Protection is therefore activated 14 windows earlier by Orbit-AWARE, before the two mean-score trajectories cross.
Safeguards are activated at W3 under Orbit-AWARE, 14 windows before the Rule-Based Drift Loop responds at W17 and before the mean-score trajectories cross.

The communication consequence of this ranking reversal is quantified in Fig.~\ref{fig:drift-delivery}(b) using \ac{COR} and the number of displaced critical observations.
% 缩写，删除一个公式
\ac{COR} measures the fraction of all critical observations in the deployment stream that are successfully delivered to the ground.
% Displacement measures the gap between the maximum number of critical observations permitted by the communication budget and the number actually delivered.
The delivery gap measures the difference between the budget-only critical-delivery ceiling and the number of critical observations delivered.
% For method $m$, these metrics are defined as
% \begin{equation}
%     \mathrm{COR}_{m}
%     =
%     \frac{D_m}{N_{\mathrm{crit}}},
%     \qquad
%     E^{\mathrm{disp}}_{m}
%     =
%     D_{\mathrm{ideal}}-D_m,
% \end{equation}
% where $D_m$ denotes the number of critical observations delivered by method $m$, $N_{\mathrm{crit}}$ is the total number of critical observations in the deployment stream, and $D_{\mathrm{ideal}}$ is the maximum critical delivery permitted by the available transmission slots.
% All methods are evaluated under the same communication budget.
% In the controlled deployment replay, $N_{\mathrm{crit}}=130$ and $D_{\mathrm{ideal}}=58$, while the same budget of 58 products, or 8.73~MB, is imposed on all methods.
The replay contains 130 critical observations.
% , and the common 8.73 MB budget permits 58 product transmissions.
The highest \ac{COR} among deployable methods is achieved by Orbit-AWARE, which delivers 40 critical observations, compared with 35, 32, and 28 under the Rule-Based Drift Loop, Static Filter, and Fixed Exploration, respectively.
Relative to the Rule-Based Drift Loop, critical delivery increases by 14.3\%, COR rises from 26.9\% to 30.8\%, and displacement decreases from 23 observations to 18.
The label-informed Oracle delivers 54 critical observations, corresponding to a COR of 41.5\%.
Early protection and subsequent evidence selection thus direct more of the available downlink capacity toward mission-critical observations.

% Together, the two panels in Fig.~\ref{fig:drift-delivery} demonstrate how ranking degradation is translated into communication-level evidence loss.
% Because the same 58-product budget is imposed on every method, the performance differences reflect  response timing and evidence selection under an identical link budget.
% % More mission-critical observations are kept visible to the ground under Orbit-AWARE's early-protection and evidence-selection policy,
% % which also provide evidence for model adaptation and candidate validation.
% Orbit-AWARE preserves ground visibility of mission-critical observations while accumulating evidence for adaptation and candidate validation.
% Early protection and subsequent evidence selection thus support both immediate task delivery and model recovery within the same downlink budget.
Under the common budget, Fig.~\ref{fig:drift-delivery} links ranking degradation to evidence loss and shows how early protection and subsequent selection preserve critical delivery while accumulating evidence for adaptation and candidate validation.

\subsection{Trusted Recovery and Sustainable Utility}
% Required packages: booktabs, array, makecell, tabularx
\begin{table*}[b]
\centering
\caption{Evidence-conditioned candidate admission in the Orbit-AWARE Trust Plane.}
\label{tab:trusted-admission}
\renewcommand{\arraystretch}{1.45}
\setlength{\tabcolsep}{4pt}
\small
\begin{tabularx}{0.98\textwidth}{@{}
    >{\centering\arraybackslash}m{0.11\textwidth}
    >{\centering\arraybackslash}m{0.13\textwidth}
    >{\centering\arraybackslash}m{0.18\textwidth}
    >{\centering\arraybackslash}m{0.10\textwidth}
    >{\centering\arraybackslash}m{0.18\textwidth}
    >{\centering\arraybackslash}X@{}}
\toprule
\makecell[c]{\textbf{Evidence} \\\textbf{checkpoint}} &
\makecell[c]{\textbf{Critical /} \\\textbf{noncritical counts}} &
\makecell[c]{\textbf{Admission evidence}} &
\makecell[c]{\textbf{Trust} \\\textbf{decision}} &
\makecell[c]{\textbf{Post-decision} \\\textbf{audit AUROC}} &
\makecell[c]{\textbf{Lifecycle} \\\textbf{consequence}} \\
\midrule
12 $\rightarrow$ 24 $\rightarrow$ 36 &
11/1 $\rightarrow$ 19/5 $\rightarrow$ 27/9 &
Evidence below the admission gate &
\textbf{Defer} &
Frozen: 0.631 &
Retain the verified model \\
\midrule
48 &
33 / 15 &
\makecell[c]{$\Delta A_{\mathrm{vis}}=-0.107$ \\ $< -\epsilon$} &
\textbf{Reject} &
\makecell[c]{Candidate: 0.651 \\ deployed: 0.631} &
Reject the candidate that fails admission \\
\midrule
58 &
40 / 18 &
\makecell[c]{$\Delta A_{\mathrm{vis}}=-0.033$ \\ $\geq -\epsilon$} &
\textbf{Accept} &
\makecell[c]{Candidate/deployed: \\ 0.696} &
Activate the admitted candidate \\
\bottomrule
\end{tabularx}
\vspace{1pt}

{\footnotesize\raggedright
\emph{Note:} $\Delta A_{\mathrm{vis}}$ is the candidate-minus-frozen AUROC on downlinked validation evidence, with the mission-configurable admission tolerance $\epsilon=0.05$.
Audit AUROC is used only for post-decision evaluation.\par}
\end{table*}

% Table~\ref{tab:trusted-admission} examines how the accumulated evidence governs whether model admission is deferred, rejected, or accepted.
Evidence-conditioned decisions to defer, reject, or accept candidate models are summarized in Table~\ref{tab:trusted-admission}.
For an evidence checkpoint containing $n$ observations, the visible performance change is defined as
\begin{equation}
    \Delta A_{\mathrm{vis}}(n)
    =
    A_{\mathrm{vis}}^{\mathrm{cand}}(n)
    -
    A_{\mathrm{vis}}^{\mathrm{frozen}}(n),
\end{equation}
where $A_{\mathrm{vis}}^{\mathrm{cand}}$ and $A_{\mathrm{vis}}^{\mathrm{frozen}}$ denote the candidate and frozen-model \ac{AUROC} values measured on downlinked validation evidence.
Candidate evaluation is deferred until the minimum evidence requirement $n_{\min}=48$ is reached.
% while the mission-configurable non-inferiority margin is set to $\epsilon=0.05$,
% allowing at most a five-percentage-point decrease invisible \ac{AUROC} before the candidate is rejected.
The visible-performance criterion requires $\Delta A_{\mathrm{vis}}(n) \geq -\epsilon$, where $\epsilon$ is the admissible \ac{AUROC} decrease.
Here, $\epsilon = 0.05$.
% At $n=48$, the first candidate yields $\Delta A_{\mathrm{vis}}=-0.107<-\epsilon$ and is rejected.
% The verified model remains in service with an audit \ac{AUROC} of 0.631.
% After the evidence set grows to 58 observations, the ground-side adapter produces a new candidate with $\Delta A_{\mathrm{vis}}=-0.033\geq-\epsilon$.
% This candidate is admitted and increases the deployed audit \ac{AUROC} to 0.696.
% The audit \ac{AUROC} is computed only after each decision and remains unavailable to the admission policy.
% These decisions illustrate the separation between producing a model update and restoring its data-disposition authority.
At $n=48$, the first candidate is rejected because its visible performance change falls below the admission tolerance, and the verified model remains in service.
At $n=58$, the new candidate satisfies the admission criterion and raises the deployed audit \ac{AUROC} to 0.696.
These decisions show how accumulated evidence governs the restoration of data-disposition authority.

\begin{figure}[tb]
    \centering
    \includegraphics[width=0.48\textwidth]{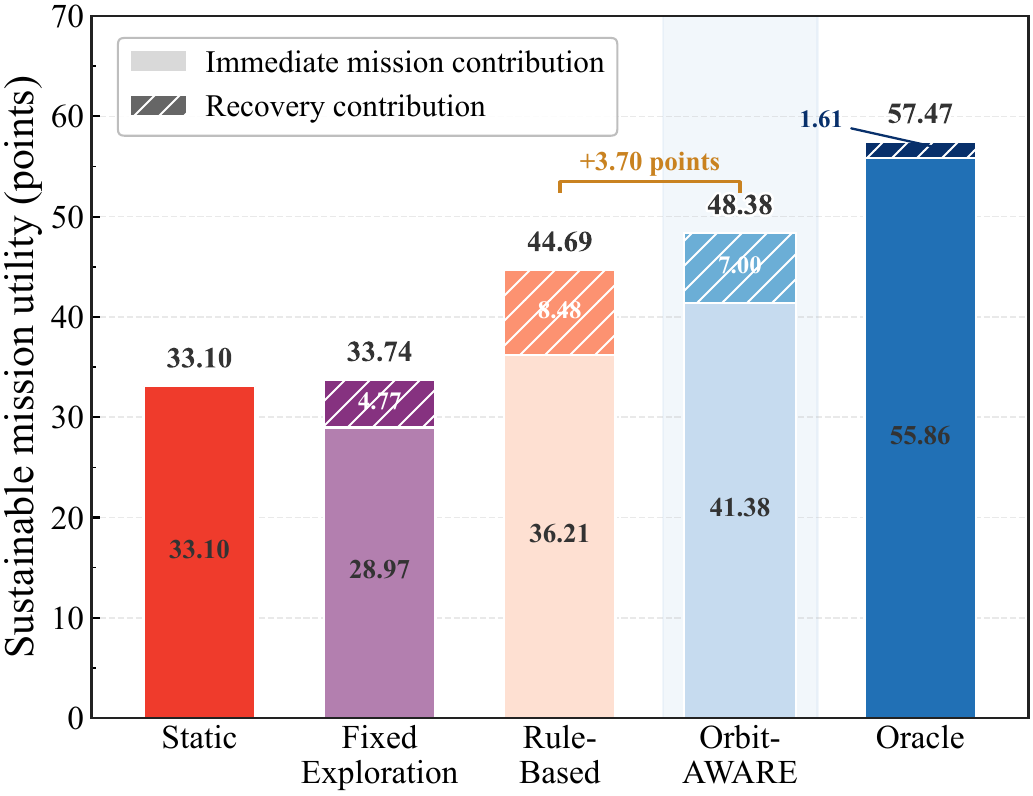}
    % \caption{Illustration of SMU under 10\% downlink budget.}
    \caption{Sustainable mission utility under the common downlink budget, decomposed into immediate mission and model-recovery contributions.}
    \label{fig:smu}
\end{figure}

% The system-level outcome of the complete lifecycle is reported in Fig.~\ref{fig:smu} using \ac{SMU}, which combines immediate evidence delivery, model recovery, and communication-budget compliance:
To capture the system-level outcome of the complete lifecycle, we define \ac{SMU}, 
which combines immediate critical-evidence
delivery, normalized model recovery, and communication-budget compliance:
% \begin{equation}
% \mathrm{SMU}
% =
% 100\left[
% \alpha\frac{D}{D_{\mathrm{ideal}}}
% +
% (1-\alpha)
% \frac{
% \max\left(0,A^{\mathrm{dep}}-A^{\mathrm{frozen}}\right)
% }{
% 1-A^{\mathrm{frozen}}
% }
% \right]\eta_B,
% \end{equation}
\begin{equation}
\mathrm{SMU}
=100\eta_B\!\left[
\alpha\frac{D}{D_{\mathrm{ideal}}}
+(1-\alpha)
\frac{\max\!\left(0,A^{\mathrm{dep}}-A^{\mathrm{frozen}}\right)}
{1-A^{\mathrm{frozen}}}
\right],
\end{equation}
where $D$ is the number of delivered critical observations,
$D_{\mathrm{ideal}}$ is the maximum critical delivery permitted by the communication budget, 
$A^{\mathrm{dep}}$ and $A^{\mathrm{frozen}}$ are the audit \ac{AUROC} values of the deployed and frozen models, respectively,
$\alpha\in[0,1]$ is the mission-immediacy weight and $\eta_B$ is the budget-compliance factor.
% determined by the permitted traffic $B_{\mathrm{limit}}$ and consumed traffic $B_{\mathrm{used}}$.
% In this emergency \ac{EO} setting, $\alpha=0.6$ assigns greater weight to immediate critical-observation delivery while retaining a 40\% contribution from normalized model recovery.
Here, $\alpha=0.6$ assigns weights of 0.6 and 0.4 to immediate critical-observation delivery and normalized model recovery, respectively, prioritizing timely disaster information.
% The budget-compliance factor is $\eta_B=\min(1,B_{\mathrm{limit}}/B_{\mathrm{used}})=1$, 
% since all methods use the common 8.73 MB budget.
The budget-compliance factor is $\eta_B=\min(1,B_{\mathrm{limit}}/B_{\mathrm{used}})$, 
where $B_{\mathrm{limit}}$ and $B_{\mathrm{used}}$ denote the available and consumed communication budgets, respectively.
It equals one for all methods, which remain within the common 8.73 MB budget.
% In the current emergency \ac{EO} setting, $\alpha=0.6$ is adopted to moderately prioritize timely critical-observation delivery while retaining substantial weight for model recovery, while the budget-compliance factor $\eta_B=\min(1,B_{\mathrm{limit}}/B_{\mathrm{used}})=1$ because every method uses the same 8.73~MB budget.
The resulting comparison is illustrated in
Fig.~\ref{fig:smu}, 
where the highest deployable \ac{SMU} of 48.38 points is obtained by Orbit-AWARE, followed by the Rule-Based Drift Loop, Fixed Exploration, and Static Filter with 44.69, 33.74, and 33.10 points, respectively.
% Relative to the Rule-Based Drift Loop, 
% % with 8.48 points compared with 7.00 points, 
% the immediate mission contribution is increased from 36.21 to 41.38 points by Orbit-AWARE.
% This delivery gain offsets the difference in recovery contribution, yielding a net \ac{SMU} improvement of 3.70 points.
% % producing a net gain of 3.70 points.
Relative to the Rule-Based Drift Loop, Orbit-AWARE raises the immediate mission contribution from 36.21 to 41.38 points and the total \ac{SMU} from 44.69 to 48.38 points.
This improvement shows how coordinated evidence protection, downlink selection, and governed recovery translate into greater lifecycle-level mission utility.
The mission-oriented Oracle reaches 57.47 points,
providing an idealized reference for label-informed observation selection under the same communication budget.
% The resulting decomposition shows that the advantage of Orbit-AWARE is produced by lifecycle-level coordination rather than dominance in every isolated component.
% The highest \ac{SMU} among the deployable methods is obtained by Orbit-AWARE through the combined effects of early protection, evidence selection, and controlled model admission.
% Together, the delivery results and admission decisions demonstrate how lifecycle coordination sustains mission utility under a fixed communication budget.
The delivery gains in Fig.~\ref{fig:drift-delivery} and the admission decisions in Table~\ref{tab:trusted-admission} demonstrate complementary lifecycle benefits: preserving mission-critical evidence and governing model reactivation.
Their joint mission-level outcome is captured by the higher \ac{SMU} under the fixed downlink budget.
% By converting early drift awareness into protected evidence delivery and subsequently into governed model reactivation, 
% Orbit-AWARE achieves the highest sustained mission utility among all deployable methods, demonstrating the effectiveness of coordinating the complete onboard \ac{AI} lifecycle rather than optimizing any isolated stage.

\section{Toward Self-Recovering Onboard Intelligence}
% 未来展望
Future research directions are outlined across the following three dimensions.

\subsection{From Drift Detection to Reliability Awareness}
% Sustaining onboard \ac{AI} models throughout their operational lifetime requires timely assessment of whether changes in observations threaten task performance or the evidence needed for recovery.
% Future monitoring should therefore estimate how observed changes affect mission delivery and model recoverability.
% A distribution shift may be harmless to one task yet disrupt the ranking used by another.
Sustainable onboard \ac{AI} requires monitoring that assesses how changing observations affect mission delivery and the evidence needed for recovery.
Task-specific reliability models are needed to connect unlabeled monitoring signals with the consequences of filtering decisions.
Delayed ground labels and independently retained observations can calibrate these estimates as operating conditions evolve.
Intervention policies should account for both the cost of unnecessary protection and the irreversible loss caused by delayed action, allowing protection and verification resources to follow the estimated mission risk.

\subsection{Native Agentic Intelligence in Orbit}
% As operational knowledge accumulates, diagnosis, evidence selection, and recovery planning must evolve throughout an onboard \ac{AI} model's lifecycle.
% Onboard \acp{LLM} and \acp{VLM} could support agents that retrieve and reuse operational knowledge across missions.
As operational knowledge accumulates, onboard \acp{LLM} and \acp{VLM} could help agents reuse experience across missions to improve diagnosis, evidence selection, and recovery planning.
Their memory should associate prior events with observation conditions, model versions, tool results, and recovery outcomes, allowing relevant experience to inform new decisions.
Coordination between onboard and ground agents could distribute diagnosis and planning according to contact availability and computational resources.
Evaluation should examine whether memory and tool use improve evidence selection and recovery decisions under unfamiliar conditions.
Structured action interfaces and independent authorization remain essential as agents acquire greater planning autonomy.

\subsection{Evidence-Centric Space Networking}
Communication resources must support both immediate mission delivery and subsequent model recovery throughout the operational life of onboard \ac{AI} models.
Extending this principle across a constellation raises networking questions about where evidence should be retained, how it should be represented, and which node should receive it.
Routing and scheduling could account for expected diagnostic and recovery value alongside delivery deadlines, evidence age, and transmission cost.
Complementary observations from different satellites may resolve ambiguities that repeated observations from one platform cannot, provided their sensing conditions and provenance are preserved.
Shared benchmarks should couple changing observation domains, selective data visibility, intermittent contacts, and model recovery to evaluate how network decisions sustain intelligence throughout a mission.

\section{Conclusion}
\label{sec:conclusion}
As space computing expands autonomous onboard data handling, this article examined drift-induced evidence displacement as a lifecycle challenge linking unreliable filtering to mission-critical observation loss and reduced recovery evidence.
% Representative approaches were compared to identify the requirements for sustained lifecycle management.
% Orbit-AWARE was then introduced to coordinate continuous data processing, event-triggered agentic orchestration, and deterministic governance through cooperative Data, Cognitive, and Trust Planes.
% Its safeguard-first lifecycle connects early evidence protection with diagnosis, evidence exchange, model adaptation, and verified restoration of data-disposition authority.
The comparison of representative approaches established requirements for sustainable \ac{LCM}. Orbit-AWARE translates these requirements into a safeguard-first architecture that coordinates evidence preservation, ground-assisted adaptation, and verified model reactivation across the Data, Cognitive, and Trust Planes. In the xBD-based case study, improved critical-observation delivery and controlled candidate admission yielded the highest \ac{SMU} among the deployable methods under the same downlink budget,  providing a basis for extending validation to on-orbit agent operation and constellation-scale coordination. 
% In addition, we highlight several future research directions, including task-aware reliability assessment, onboard agents with operational memory and tool-use capabilities, and evidence-centric space networking. 
Future priorities include task-aware reliability assessment, memory- and tool-enabled onboard agents, and evidence-centric space networking.
The lifecycle perspective developed in this article provides architectural principles for integrating communication, learning, and trusted autonomy, shifting the focus from model deployment to sustainable, self-recovering onboard intelligence.

\section*{Acknowledgement}
This work was supported by Fundamental and Interdisciplinary Disciplines Breakthrough Plan of the Ministry of Education of China under Grant JYB2025XDXM119; National Natural Science Foundation of China under Grants 62595741, 62595745, 62595740, 62401070, and 62321001; Natural Science Foundation of Beijing under Grant L232002 and L242012; The Open Project of Anhui Provincial Key Laboratory of Multimodal Cognitive Computation, Anhui University, No. MMC202403.

% reference
\bibliographystyle{IEEEtran}
\bibliography{reference}

\begin{IEEEbiographynophoto}{Yunxin Geng} (Graduate Student Member, IEEE) received the M.E. degree from Beijing University of Posts and Telecommunications, China, in 2023, where he is currently pursuing his Ph.D. degree. He is an engineer with Shanghai Satellite Network Research Institute Co., Ltd. His research interests include space-ground integrated networks, space computing, constellation management and deep learning.
\end{IEEEbiographynophoto}

\begin{IEEEbiographynophoto}{Kaige Zhu} received the Ph.D. degree in electrical and communication engineering from Beijing Jiaotong University (BJTU), Beijing, China, in 2023. He is an engineer with Shanghai Satellite Network Research Institute Co., Ltd. His research interests include edge computing and machine learning techniques.
\end{IEEEbiographynophoto}

\begin{IEEEbiographynophoto}{Tongtong Shu} received the M.E. degree from Beihang University, Beijing, China, in 2022. He is an engineer with China Satellite Network Group Co., Ltd. His research interests include satellite communication and space-ground integrated networks.
\end{IEEEbiographynophoto}

\begin{IEEEbiographynophoto}{Zhan Gao} received the Ph.D. degree in Computer Application Technology from Beijing Jiaotong University (BJTU), Beijing, China, in 2010. He is a researcher at Shanghai Satellite Network Research Institute Co., Ltd., whose research focuses on computing-network convergence and machine learning.
\end{IEEEbiographynophoto}

\begin{IEEEbiographynophoto}{Yanxiao Li} received the M.S. degree from Civil Aviation University of China in 2017. He is currently pursuing the Ph.D. degree at Beijing University of Posts and Telecommunications, China. He is an engineer with China Satellite Network Group Co., Ltd. His research interests include space-ground integrated networks, satellite network routing, and computing-aware routing.
\end{IEEEbiographynophoto}

\begin{IEEEbiographynophoto}{Yan Liu} received the B.Eng., M.S. and Ph.D. degrees from Shandong University, Jinan, China, in 2007, Beihang University, Beijing, China, in 2010, and Tsinghua University, Beijing, China, in 2018, respectively. He is a senior expert at Shanghai Satellite Network Research Institute Co., Ltd. His research interests include computing-network convergence, artificial intelligence, machine learning, wireless communication, signal processing, information theory and statistics, and edge computing.  
\end{IEEEbiographynophoto}

\begin{IEEEbiographynophoto}{Weiwei Jiang}(Senior Member, IEEE) received the B.Sc. and Ph.D. degrees from the Department of Electronic Engineering, Tsinghua University, Beijing, China, in 2013 and 2018, respectively. He is currently an associate professor with the School of Information and Communication Engineering, Beijing University of Posts and Telecommunications, and Key Laboratory of Universal Wireless Communications, Ministry of Education. His current research interests include artificial intelligence, machine learning, big data, wireless communication and edge computing.
\end{IEEEbiographynophoto}

\begin{IEEEbiographynophoto}{Zhiqing Wei} (Member, IEEE) received the B.E. and Ph.D. degrees from Beijing University of Posts and Telecommunications (BUPT), Beijing, China, in 2010 and 2015, respectively. He is a Professor
with BUPT. His research interest is the performance analysis and optimization of intelligent machine networks. He was granted the
Exemplary Reviewer of IEEE WIRELESS COMMUNICATIONS LETTERS in 2017, the Best Paper Award of WCSP 2018 and WCSP 2022. He was the Registration Co-Chair of IEEE/CIC ICCC 2018, the publication Co-Chair of IEEE/CIC ICCC 2019 and IEEE/CIC ICCC 2020.
\end{IEEEbiographynophoto}

\begin{IEEEbiographynophoto}{Zhiyong Feng} (Senior Member, IEEE) received the B.E., M.E., and Ph.D. degrees from Beijing University of Posts and Telecommunications (BUPT), Beijing, China. She is a Professor with BUPT, and the Director of the Key Laboratory of the Universal Wireless Communications, Ministry of Education, China. She is a Vice Chair of the Information and Communication Test Committee of the Chinese Institute of Communications (CIC). Her main research interests include wireless network architecture design and radio resource management in 5th generation mobile networks (5G), spectrum sensing and dynamic spectrum management in cognitive wireless networks, and universal signal detection and identification.
\end{IEEEbiographynophoto}

\vfill

\end{document}